\documentclass[prd,aps,twocolumn]{revtex4}
\usepackage{graphicx,url,color,amsmath,amssymb}

\renewcommand{\vec}[1]{\boldsymbol{#1}}

\def\gr{$\gamma$-ray}
\def\lsim{\raise0.3ex\hbox{$\;<$\kern-0.75em\raise-1.1ex\hbox{$\sim\;$}}}
\def\gsim{\raise0.3ex\hbox{$\;>$\kern-0.75em\raise-1.1ex\hbox{$\sim\;$}}}

\def\R{{\cal R}}
\def\be{\begin{equation}}
\def\ee{\end{equation}}
\def\ba{\begin{align}}
\def\ea{\end{align}}

\begin{document}

\title{Cosmic ray signatures of a 2--3\,Myr old local supernova}

\author{M.~Kachelrie\ss$^{1}$}
\author{A.~Neronov$^{2}$}
\author{D.~V.~Semikoz$^{3,4}$}
\affiliation{$^1$Institutt for fysikk, NTNU, Trondheim, Norway}
\affiliation{$^2$Astronomy Department, University of Geneva,
Ch. d'Ecogia 16, Versoix, 1290, Switzerland}
\affiliation{$^{3}$APC, Universite Paris Diderot, CNRS/IN2P3, CEA/IRFU,
Observatoire de Paris, Sorbonne Paris Cite, 119 75205 Paris, France}
\affiliation{$^{4}$National Research Nuclear University MEPHI (Moscow Engineering Physics Institute), Kashirskoe highway 31, 115409 Moscow, Russia}

\begin{abstract}
The supernova explosion which deposited $^{60}$Fe isotopes on Earth 
2--3 million years ago should have also produced cosmic rays which contribute 
to the locally observed cosmic ray flux. We show that the contribution of 
this ``local source'' causes the ``anomalies'' observed in the positron and 
antiproton fluxes and explains why their spectral shapes agree with that of the proton flux.  At the same time, this  local source component accounts for 
the difference in the 
slopes of the spectra of cosmic ray nuclei as the result of the slightly 
varying relative importance of the ``local'' and the average component for 
distinct CR nuclei. Such a  ``local supernova'' model for the spectra of 
nuclei can 
be tested via a combined measurement of the energy dependence of the 
boron-to-carbon (primary-to-secondary cosmic rays) ratio and of the 
antiproton spectrum: While the antiproton spectrum is predicted to extend 
approximately as a power law  into the TeV range without any softening break,
the B/C ratio is expected to show a "plateau" at a level  fixed by the observed 
positron excess in the 30--300\,GeV range.  We discuss the 
observability of such a plateau with  dedicated experiments for the 
measurement  of the cosmic ray composition in the 10\,TeV energy range 
(NUCLEON, ISS-CREAM).   
\end{abstract}


\maketitle

\section{Introduction}

\begin{figure}
\includegraphics[width=\columnwidth]{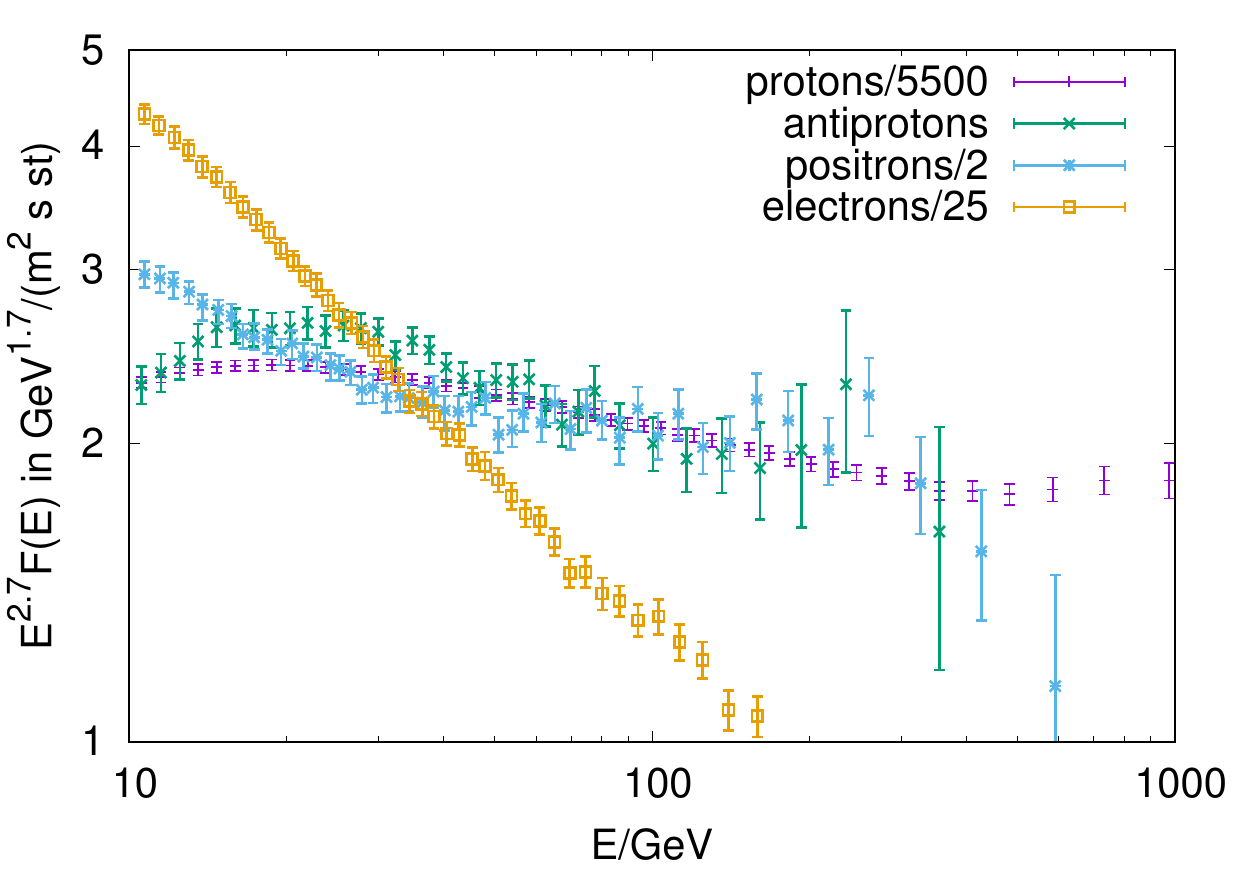}
\caption{Proton flux (rescaled by 1/5500), electron flux  (rescaled by 1/25),
positron flux (rescaled by 1/2)  and the antiproton flux  measured by AMS-02;
all multiplied by $E^{2.7}$.}
\label{fig:sec}
\end{figure}

The spectra of cosmic rays (CR) measured locally possess a number of 
puzzling features which are not well explained assuming a smooth distribution 
of CR sources, as it is often done in the conventional diffusion 
approach~\cite{SM}. In particular, 
\begin{enumerate}
\item[$(i)$] 
the slopes of the spectra $dN/d\R$ of different nuclei slightly 
differ~\cite{PAMELA_pHe,AMS02_p,AMS02_He}, although both acceleration 
and diffusion depend only on rigidity $\R=cp/Z$;
\item[$(ii)$]  
the spectra of different nuclei show a softening above $\simeq 10$\,GV, 
followed by a hardening above several hundred GV~\cite{PAMELA_pHe,AMS02_p,AMS02_He,CREAM,CREAM_nuclei,CREAMIII};
\item[$(iii)$]  
the spectrum of positrons shows an excess above the energy $E\simeq 30$\,GeV 
compared to the naive expectation from secondary production during propagation 
of CR nuclei in the interstellar medium~\cite{PAMELA_pos,AMS02_pos};
\item[$(iv)$]  
similarly, the spectrum of antiprotons above $\simeq 100$\,GeV is 
harder than expected from secondary production~\cite{PAMELA_pbar,AMS02_pbar};
\item[$(v)$]  
the positron and antiproton fluxes above 100\,GeV repeat the 
spectral shape of the proton flux; in contrast, the electron flux
is considerably steeper~\cite{AMS02_pos,AMS02_pbar}, as shown in 
Fig.~\ref{fig:sec};
\item[$(vi)$]  
the amplitude $\delta$ of the dipole anisotropy 
of the CR flux has a nearly constant level in the 1--30\,TV rigidity range, 
contrary to the expected increase $\delta\propto D(\R)$ proportional to the 
rigidity dependence of the diffusion coefficient $D(\R)$~\cite{aniso}. 
\end{enumerate}
A ``patchwork'' of different solutions has been suggested to explain subsets
of the features $(i)$--$(vi)$. Additional sources of primary positrons  such 
as pulsars~\cite{pulsars}, or annihilations or decays of dark matter~\cite{dm} 
may 
explain the observed positron excess $(iii)$. Uncertainties of the antiproton 
production cross-section and of CR propagation models are invoked to relieve 
the problem $(iv)$~\cite{antip}. Peculiarities of the propagation of 
CRs~\cite{prop} through the interstellar medium, e.g.\ a non-factorizable
rigidity and space  dependence of the diffusion coefficient, have been 
considered as solutions for problem $(ii)$.

In this work, we address the question if a single phenomenon can explain 
the entire set of features $(i)$--$(vi)$.
Perhaps the two most important clues how to solve these puzzles at one stroke
are the problems $(v)$ and $(vi)$: the nearly scale invariance of hadronic 
interactions implies that the secondary fluxes produced in interactions of 
CRs on interstellar gas have a shape similar to the primary flux, if the 
grammage CRs cross is energy independent. Such an energy independence of
the grammage is achieved if a young source contributes significantly 
to the observed local CR flux~\cite{PRL}. 
Moreover, the ratio $R(E)=F_{e^+}(E)/F_{\bar p}(E)$ of the positron
and antiproton fluxes is then mainly predicted by the properties of hadronic 
interactions,  being proportional to the ratio of their $Z$ factors,  
$R=F_{e^+}/F_{\bar p}\propto Z_{e^+}/Z_{\bar p}$~\cite{PRL,lipari}. 
As we showed in our previous {\em Letter\/}~\cite{PRL},  
the observed values of $R$ lie within the range
expected in this scenario. In contrast, for other primary 
sources of antimatter as dark matter annihilations or pulsars one expects 
naturally a much larger value of $R$. Recall also that the energy-dependence 
$X\propto \R^{-\beta}$ with $\beta=0.3-0.5$ expected in standard diffusion 
models would lead to a similar decrease of the secondary/proton ratios,
which is not seen in Fig.~\ref{fig:sec}. 
Finally, the observed plateau in the CR dipole anisotropy $\delta$ in
the energy range $2-20$\,TeV can be also 
explained in the ``local source'' model, because the anisotropy $\delta$ 
of a single CR source is rigidity independent~\cite{savchenko}.
The decrease of the measured dipole anisotropy at lower energies,
$E\lsim 2$\, TeV is naturally explained by an off-set of the
Earth with respect to magnetic field line going through
the source~\cite{savchenko}.
Aim of the present work is to extend the {\em Letters}~\cite{PRL,savchenko} 
and to show that our scenario explains also naturally the remaining puzzles
connected to the spectra of CR nuclei. Moreover, we have to ensure
that the low-energy cutoff for the local CR flux does
not spoil our succesfull explanation of the secondary fluxes obtained in
Ref.~\cite{PRL}.

Let us now sketch why a single CR source may give a dominant
contribution to the local CR flux in the TV range. 
In the conventional diffusion approach one uses a {\em smooth\/} distribution 
of CR sources and searches for a {\em stationary} solution of the coupled 
cascade equations. These two approximations seem natural in the GV--TV 
rigidity range if one adopts the isotropic diffusion of CRs with diffusion 
coefficient $D\simeq 10^{28}\left( \R/5 {\rm GV}\right)^\beta$\,cm$^2$/s and
$\beta\simeq 0.3-0.5$. If one moreover assumes that the CR sources are 
supernovae (SN) depositing  some $\sim 10^{50}$\,erg  every $T\sim 30$~yr 
in the form of CRs which  escape with the characteristic time scale 
$T_{\rm esc}\simeq {\rm few}\times 10^7{\rm yr}\left(\R/5\, {\rm GV}\right)^{-\beta}$, then the flux from some $10^{4}$ sources 
accumulates at low rigidities, forming a ``sea'' of Galactic CRs.
As a result, Ref.~\cite{myriad} concluded that the dominance of a
single source corresponds to a rare fluctuations and is therefore
extremly unlikely in the isotropic diffusion picture.

The approximation of continuous injection might, however, not be valid 
if CRs propagate strongly anisotropically. Such an anisotropy may appear if
the turbulent field at the considered scale does not dominate over the 
ordered component, or if the turbulent field itself is anisotropic.
As the authors of Ref.~\cite{GMF} stressed recently, anisotropic CR
propagation is necessary, because otherwise CRs overproduce
secondary nuclei like boron for any reasonable values of the strength of
the turbulent field. As a result, they concluded that the number of sources
contributing to the local CR flux is reduced by a factor ${\cal O}(100)$
relative to isotropic CR diffusion. For the special case of the
Jansson--Farrar model~\cite{JF} for the Galactic magnetic field, the authors
of Refs.~\cite{GKS14,GKS15} obtained a satisfactory description of all data
on Galactic primary CRs above $\simeq 200$\,GV, after the strength of the
turbulent field was reduced. As a result of this
reduction, CRs propagation becomes strongly anisotropic and the diffusion 
coefficient perpendicular to the ordered field can be between two and three 
orders of magnitude smaller 
than the parallel one, $D_\bot\ll D_{||}$. Therefore, CRs spread perpendicular
to the magnetic field line just $\sim 100$\,pc on the time scale of 
escape, compared to $\sim 1$\,kpc in the isotropic diffusion model. The 
smaller volume occupied by CRs from a single source leads to a smaller number 
of sources contributing substantially to the local flux, with only 
$\sim 10^2$ sources at $\R\sim 10$\,GV and about $\sim 10$~most recent SNe in 
the TV range.

Valuable information on recent local SNe comes from the discovery of 
radionuclides like $^{60}$Fe in the deep ocean crust~\cite{fe60}. In particular,
 Ref.~\cite{Sc17} derived the possible time sequence and locations of SN
explosions from the mass spectrum of the perished members of certain nearby 
moving stellar groups. They found that 10--20 SNe are 
responsible for the formation of the Local Bubble, while the most recent 
ones happened 2--3\,Myr ago and $\sim 100$\,pc away. According to 
Ref.~\cite{Sc17}, this SN, or a combined action of several SNe,  
may be responsible for the $^{60}$Fe deposition on Earth.

In what follows we show that the contribution to the flux of CR nuclei 
from a several Myr old supernova also explains the features ($i$) and ($ii$). 
Thus, the "local source" model provides a single explanation for the entire 
set of peculiar features ($i$)--($vi$) in the GeV--TeV CR flux. We show that 
within such a model, measuring the parameters of the features ($i$)--($v$) 
provides information on the most recent local SN. 
We also point out that the local source model is testable through its 
imprint on the boron-to-carbon (B/C) ratio. This ratio is predicted to reach a 
"plateau" in the energy range 1--10\,TeV where the local SN
component provides the strongest contribution to the CR flux.  
The level of the plateau is not a free parameter, but is fixed by 
the measurement of the positron flux at several hundred GeV. 
The (non-) detection of this plateau in the B/C ratio could, therefore, be 
used to verify (falsify) the model.  
More generally, our model relies on the assumption that relatively few sources
contribute to the CR flux. Therefore the model predicts that the CR flux
has several breaks which show up in secondary-to-primary ratios as B/C
as step-like features.

\section{Cosmic ray primary nuclei}

Measurements of the spectra of cosmic ray nuclei by PAMELA, CREAM and AMS-02 
have firmly established that the slopes of the spectra of individual nuclei 
are different: For instance, the helium spectrum above 10\,GV becomes 
increasingly harder with increasing $\R$,  compared to the proton spectrum. 
Moreover, the spectra of most of the nuclei show deviations from a single 
power law. One of the slope changes in these power-laws  is found in the 
$\R\sim 200-300$\,GV range where the proton spectrum 
$dN_p/d\R\propto \R^{-\gamma_p}$ hardens from $\gamma_p\simeq 2.8$--2.85 to 
$\gamma_p\simeq 2.6$--2.65. Another example is the helium spectrum  
$dN_{\rm He}/d\R\propto \R^{-\gamma_{\rm He}}$ which hardens from 
$\gamma_{\rm He}\simeq 2.78$ to $\gamma_{\rm He}\simeq 2.66$.

Both the difference and the changes in the slopes can be well described 
within the following simple model which includes two "template" functions, 
$F^{(1,2)}(\R)$ for the shape of the spectra as a function of rigidity $\R$, 
summed with coefficients $C^{(1/2)}_{A}$ to fit the spectra of protons $p$ and 
of nuclei with mass number $A$,
\begin{eqnarray}
\label{eq:fit}
F_p(\R)&=& C^{(1)}_{p}F^{(1)}(\R) + C^{(2)}_{p}F^{(2)}(\R), \nonumber\\
F_{A}(\R)&=& C^{(1)}_{A}F^{(1)}(\R) + C^{(2)}_{A}F^{(2)}(\R) .
\end{eqnarray}

The low-energy component $F^{(1)}(\R)$ represents the average CR flux in 
the local interstellar medium, as derived from \gr\ observations of nearby 
molecular clouds~\cite{gouldbelt}. It is a broken power law 
$F^{(1)}(\R)\propto \R^{-\gamma_{1,2}}$ with the slope changing from $\gamma_1=2.4$
to $\gamma_2=3$ at the rigidity $\R_{\rm br}=20$\,GV. This form of the average 
spectrum is also consistent with the combined AMS-02, PAMELA and Voyager~1 
measurements of the local  interstellar CR spectrum~\cite{voyager}.

\begin{figure}
\includegraphics[width=1.\columnwidth,angle=0]{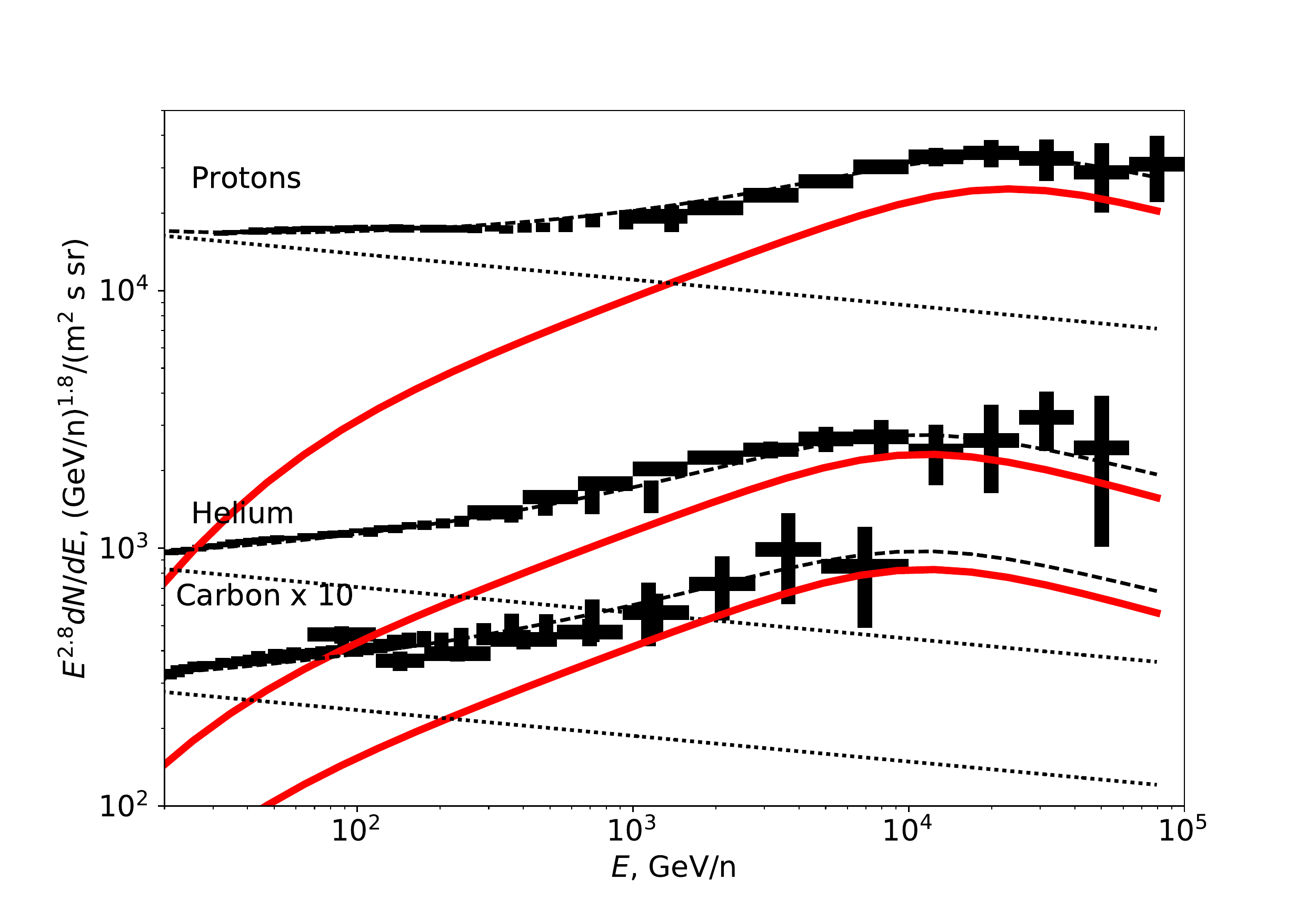}
\caption{The flux of CR protons and carbon measured by AMS-02 and 
CREAM-III as function of energy/nucleon shown together with  a 
two-component model consisting of the average CR spectrum (dotted lines) 
and the local source contribution (solid red  lines).
\label{fig:nuclei}}
\end{figure}

In addition to the average CR flux, we introduce the "local source" 
component $F^{(2)}(\R)$ which is responsible for the deviations from the 
power-law extrapolation of the average flux component. We determine the 
shape of this component using two different 
approaches: A phenomenological one and a numerical method based on the
calculation of CR trajectories in the Galactic magnetic field. 
In the phenomenological approach we fix the shape of the local source 
component by fitting the combined AMS-02 and CREAM-III spectra of CR protons, 
as shown in Fig.~\ref{fig:nuclei}.  The resulting additional component of 
the spectrum is shown by the red solid line in Fig.~\ref{fig:nuclei}.
One can see that this component has a low-energy suppression.

This low-energy suppression is naturally explained within a numerical 
approach to the modelling of the CR flux from a local source. We 
calculate this flux using the  method described in Ref.~\cite{PRL}, 
see the appendix for details. Motivated
 by the results of Ref.~\cite{Sc17}, we use as distance to 
the supernova  along the magnetic field line $d_{||}=100$\,pc and an age 3\,Myr.
The Galactic magnetic field model 
adopted in the calculation is that of Ref.~\cite{JF}, where we re-scaled
the turbulent component by a factor 1/10 as described in Ref.~\cite{GKS14,GKS15}.
The maximal scale of the turbulent field modes was set to $L_{\max}= 25$\,pc 
and a Kolmogorov power spectrum was used.

\begin{figure}
\includegraphics[width=\columnwidth]{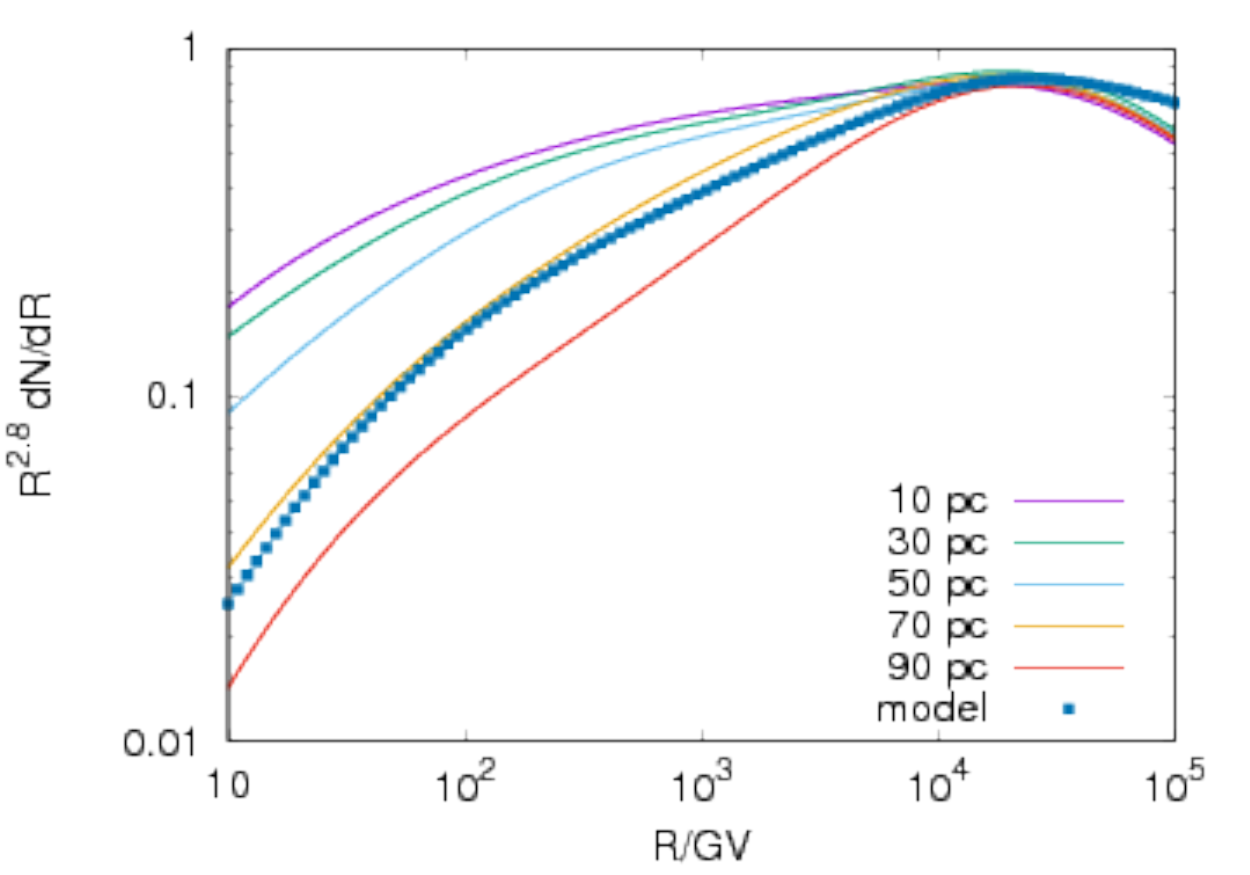}
\caption{CR rigidity spectrum for a source with $d_\|=100$\,pc and
 $d_\perp=10$, 30, 50, 70\,pc (from top to down) shown as lines compared to
the proton spectrum in the phenomenological model shown by dots.
\label{perp}}
\end{figure}

The resulting CR proton fluxes from a source with the distance
$d_\|=100$\,pc 
along the magnetic field line and a varying perpendicular distance $d_\perp$
are shown in Fig.~\ref{perp}. Since the diffusion perpendicular 
to the magnetic field is slow, a non-zero distance $d_\perp$ between the 
Solar system and the magnetic field line passing through the SN leads to a
low-energy cutoff in the locally observed CR flux from this SN.
 Comparing to the proton flux in the phenomenological model shown 
by dots to those obtained in the trajectory approach, one sees that
the model corresponds well to the case of  $d_\perp\simeq 70$\,pc. 

The energy of the low-energy cut-off is constrained by the decrease of the 
dipole anisotropy of the CR flux below several hundred GeV \cite{savchenko}.
The presence of a low-energy cut-off is also supported by the absence of an
excess flux of secondary positrons in the energy range below $\sim 30$\,GeV:
A sizable contribution  of protons from the local source in the energy 
range below several hundred GeV would produce an excess of $E<30$~GeV 
positrons~\cite{PRL}.

The propagation of CRs through the Galactic magnetic field is determined by 
the rigidity of the particles. Different nuclei with the same rigidity
diffuse identically. This means that once the shape of the local source 
component is fixed from the fit to the proton spectrum, or based on our
numerical results, the spectra of different nuclei are described by a set 
of fitting functions of the form (\ref{eq:fit}), with the coefficients 
$C^{(i)}_A$ determined by the relative abundances of different 
nuclei in the average CR flux and in the local source components. 

These relative abundances do not need to be identical for the two components. 
If one assumes that  
CRs are accelerated mostly at the forward shock of the expanding SN remnant, 
the coefficients $C^{(i)}_A$  are determined by the composition 
of the interstellar medium and the injection efficiencies of various 
elements. Since the local source exploded very likely in a star-forming 
region, deviations from the average values should be expected. Similarly, 
if a fraction of the CR flux originates from acceleration at the reverse 
shock, the  coefficients $C^{(i)}_A$ also vary depending on the 
mass of the exploded star.

Figure~\ref{fig:nuclei} shows that the two component model based on the
 proton spectrum also provides a good fit to the helium spectrum over the 
entire energy range. The local source component of the helium spectrum, 
shown by the semi-transparent red curve is shifted by a factor of two in 
energy (per nucleon), compared to the local source component of the proton 
flux. Besides the relative normalization of the average and local source 
components for protons and helium are different. The same model is applicable 
to other nuclei, and the case of the CNO group is shown in 
Fig.~\ref{fig:nuclei} too. 
Finally, the change in the slope of the CR intensity from the local source
visible at $\simeq 10$\,TeV is connected to our choice of its maximal energy.
Above this energy and up to the knee, the CR spectrum may be dominated by
another, younger source.

\section{Positrons and antiprotons}

In Ref.~\cite{PRL}, we discussed already the contribution of a local
source to the observed positron and antiproton flux. In that work
we assumed, however, that the local source and the Solar system are
connected by the same magnetic field line. In the present work this 
assumption is relaxed and, therefore, the positron and antiproton fluxes 
need to be re-calculated.

\subsection{Propagation effects for the local flux}

We have argued that the ratio of the secondary fluxes of positrons and 
antiprotons is determined by the properties of hadronic interactions, 
modulo propagation effect. As first step, 
let us look in more detail at the ratio $R=F_{e^+}(E)/F_{\bar p}(E)$ 
of the positron and antiproton flux. Neglecting propagation effects, $R$ is 
given by the ratio of the $Z$ factors of positron and antiprotons, 
respectively. The latter are defined for the secondary type $j$
via its inclusive spectrum $d\sigma_{j}(E,z_{j})/d z_{j}$, 
$z_{j}=E_{j}/E$, as
\begin{equation}
\label{Z_spec}
 Z_{j}(E_{j},\alpha) = \frac{1}{\sigma_{\rm inel}}\:\int_0^1 d z \, z^{\alpha-1}\,
 \frac{d\sigma_{j}(E_{j}/z,z)}{d z}\,,
\end{equation}
if the primary protons follow the power law $dN/dE\propto E^{-\alpha}$.
In Ref.~\cite{lipari}, this ratio was calculated  analytically in the 
asymptotic limit, i.e.\ for energies well above the threshold $E\gg E_{\rm th}$, 
as $R=1.8\pm 0.5$  for a power-law spectrum of protons with slope 
$\alpha=2.8$. We use the modified version QGSJET-IIm~\cite{QGS} presented 
in Ref.~\cite{QGSm} to calculate the positron and antiproton secondary fluxes. 
For the same set-up,  a power law with $E_{\max}\gg E$, we obtain 
$R=1.8$ ($\alpha=2.6$) and $R=1.6$ ($\alpha=2.8$). 
Using the spectrum given by Eq.~(\ref{eq:fit}) with  $E_{\max}= 1$\,PeV
as a high-energy cut-off, we find that the asymptotic 
limit represents only in a small energy range a good approximation: Below 
secondary energies $E\simeq 100$\,GeV, the threshold suppression of 
antiproton production increases $R$, while the soft positron spectra in 
forward direction lead towards $E_{\max}$ to a decrease of $R$.

\begin{figure}
\includegraphics[height=\columnwidth,angle=-90]{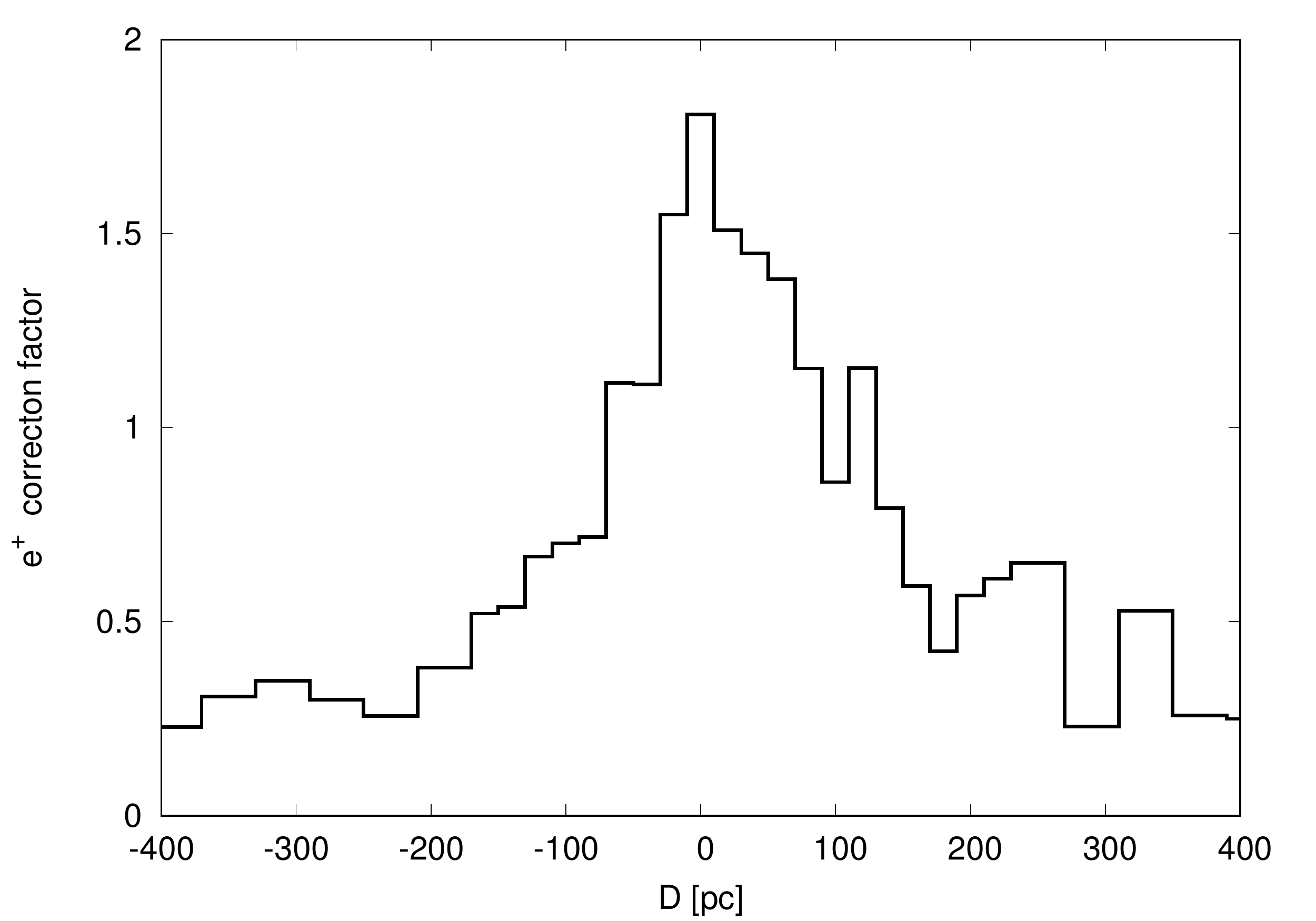}
\caption{Correction factor for positrons with energy 100\,GeV as
function of the perpendicular distance $d_\perp$ to the magnetic field line.}
\label{fig:enhancement}
\end{figure}

Let us now discuss how propagation effects modify the secondary/proton ratio.
Positrons and antiprotons have lower energies than their parent cosmic rays
and spread therefore more slowly into the interstellar medium. As a result, 
they occupy a smaller volume than their parent cosmic rays. Within this 
smaller volume, the positron and antiproton fluxes are enhanced compared to 
the reference flux which one would obtain assuming that secondaries 
propagate in the same way as the primary particles. In contrast, at larger
perpendicular distances, the secondary flux is suppressed relative to
the reference flux, because the secondaries diffuse slower.

In Ref.~\cite{PRL}, we calculated analytically a correction factor for the 
positron and antiproton flux for the particular case, when the observer is 
situated at the same magnetic field line as the source. In this case, the 
positron and the antiproton flux experience an enhancement by a factor of 
about two. Figure~\ref{fig:enhancement} shows the result of our numerical
calculations corresponding to the more generic situation when the observer 
is separated by the distance $d_\perp$ from the magnetic field line going 
through the source. Here, the positive and negative values of $d_\perp$ 
correspond to the case that the observer is displaced towards the Galactic 
Center or away, respectively.

This correction factor is calculated in following way. We divide the production
time in six intervals of $0.5$\,Myr duration. In each time interval, we assume 
that positrons are produced by protons with 20 times higher energy, which
are spatially distributed according to their numerically calculated 
trajectories. Once produced, we propagate the positrons the remaining time up 
to 3\,Myr. 
Then we obtain the positron fluxes at various perpendicular distances summing 
up the partial fluxes from all time-intervals. Dividing finally these 
fluxes by the proton flux (at the same energy), we obtain the
correction factor due to the difference in the propagation between protons and 
positrons.
From Fig.~\ref{fig:enhancement} one sees that for $d_\perp=0$  the correction
factor is close to two, as calculated analytically in Ref.~\cite{PRL}.
Increasing $d_\perp$,  the correction factor decreases, falling
below one at $d_\perp\sim 100$\,pc.

\subsection{Secondary fluxes}

\begin{figure}
\includegraphics[width=\columnwidth]{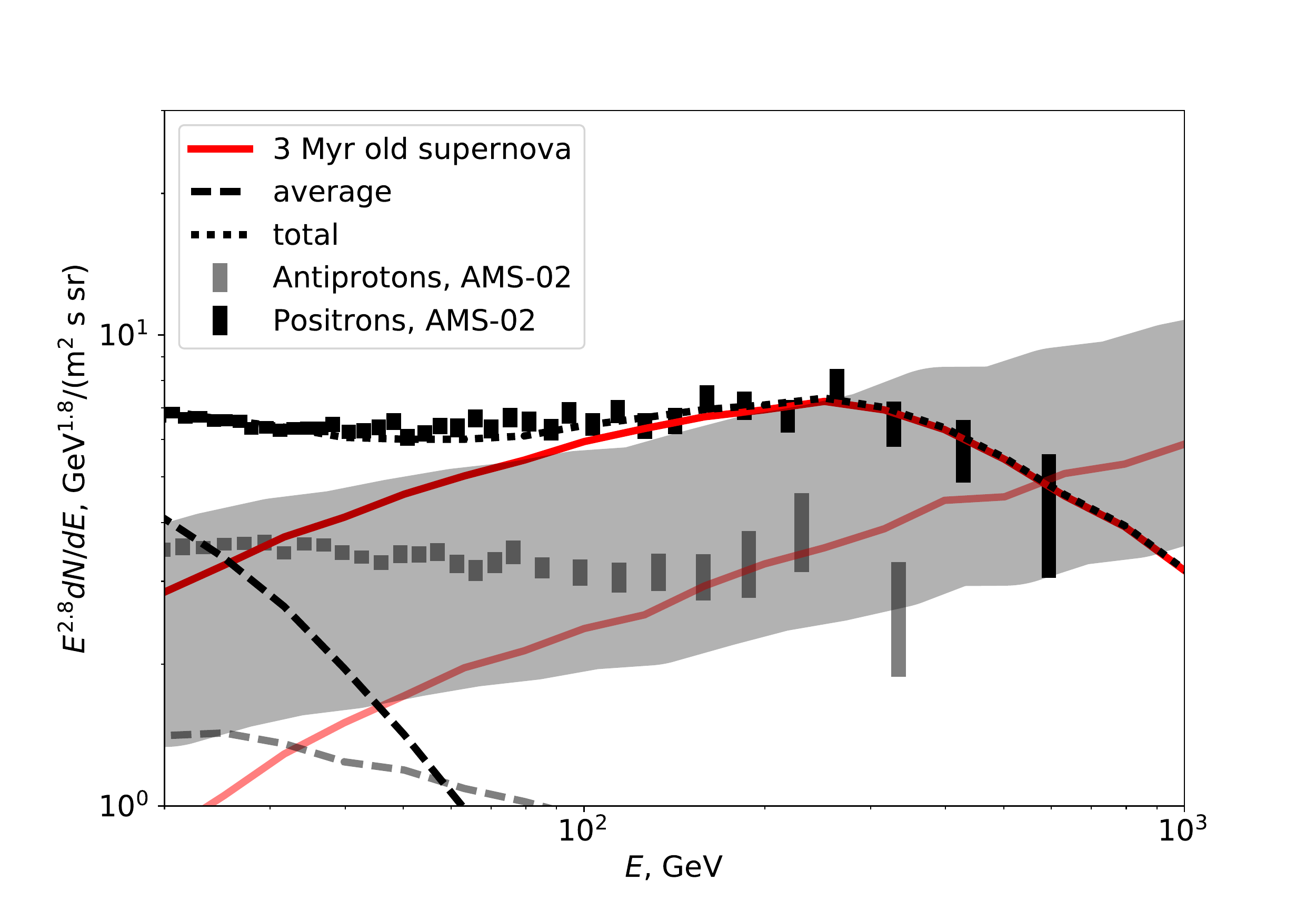}
\caption{Positron and antiproton fluxes measured by AMS-02 compared to the 
contributions for the local source (red/rose solid curves) and average (dashed black/grey) components. 
Wide grey band shows the $\pm 50\%$ uncertainty band for the sum of the 
antiproton fluxes of the two components.}
\label{fig:pos_aprot}
\end{figure}

We now turn to the resulting flux of positrons and antiprotons produced by 
CRs with the spectra $F^{(1)}(\R)$ and $F^{(2)}(\R)$. 
We obtain a reasonable fit of the measured positron flux for a range of
perpendicular distances, $d_\perp\sim \pm(50-90)$\,pc. 
In Fig.~\ref{fig:pos_aprot}
we show the positron flux for the case $d_\perp=70$\,pc, assuming that the CRs 
from the local source have traversed the column density of matter 
$X=1.2$\,g/cm$^2$, while we use $X=11.4(\R/10\rm{GV})^{-1/3}$\,g/cm$^2$ 
for the average flux of ``sea'' CRs.  
Our estimate for the grammage crossed by CRs emitted from the local source 
based on the positron flux suffers from uncertainties in the distance between 
the magnetic field lines of the source and of the observer. We use the 
reference distance 70\,pc at which the correction factor for 
the positron flux at 100\,GeV is about one. Taking the distance range 
$\pm 20$\,pc would change the correction factor by $\simeq 50\%$, 
as it is clear from Fig.~\ref{fig:enhancement}. This uncertainty propagates 
to a 50\% uncertainty in the grammage. We take this uncertainty into account 
in the analysis reported below. Note, however,  
that additional uncertainties in the Galactic magnetic field model and 
the value of $d_\perp$ exist.
As it was previously shown in 
Ref.~\cite{PRL}, the local source contribution explains the excess flux 
of CR positrons. The flux of antiprotons agrees within errors with that 
measured by AMS-02 up to 300\,GeV. 
The grammage  $X=(1.0-1.4)$\,g/cm$^2$ for CRs from the 
local source is larger than the one found in Ref.~\cite{PRL}, both
because of the used larger age of the source and the smaller proton flux
caused by the non-zero offset of the source. 
It is a factor $\simeq 2$ larger than expected for CRs propagating in a gas
density following a Nakanishi-Sofue profile~\cite{gas}.
Taking into account possible local deviations from the global gas profile 
as well as astrophysical uncertainties,  we consider this as a minor 
deviation.

At the time of publication of Ref.~\cite{PRL} the measurements of the positron 
spectrum did not extend to high enough energy to probe the effect of Compton 
cooling on the positrons injected by the local source. Meanwhile an updated 
measurement of the positron spectrum by AMS-02 has appeared~\cite{AMS02new}. 
This measurement reveals a break in the positron spectrum at 300\,GeV. Such 
a break is expected in the local source model \cite{PRL}, because positrons 
with energies around 300\,GeV loose energy via synchrotron and inverse Compton 
emission  on the time scale of 2--3\,Myr. Thus, the spectrum of positrons 
from the local source is expected to have a cooling break at about 
300\,GeV~\cite{PRL}, with the slope changing by $\Delta\alpha=1$, 
as shown in Fig.~\ref{fig:pos_aprot}. The presence of the 
cooling break is consistent with the updated AMS-02 measurement, shown by the 
black data points in the same figure. The measurement of the break energy 
determines the time elapsed since the injection of CRs by the supernova.

Contrary to positrons, antiprotons do not suffer from Compton cooling and
thus no cooling break is expected in the antiproton spectrum. The extension 
of the AMS-02 measurements to energies above 300\,GeV should reveal a nearly 
power-law continuation of the antiproton flux.

\section{Electrons}

\begin{figure}
\includegraphics[width=\columnwidth]{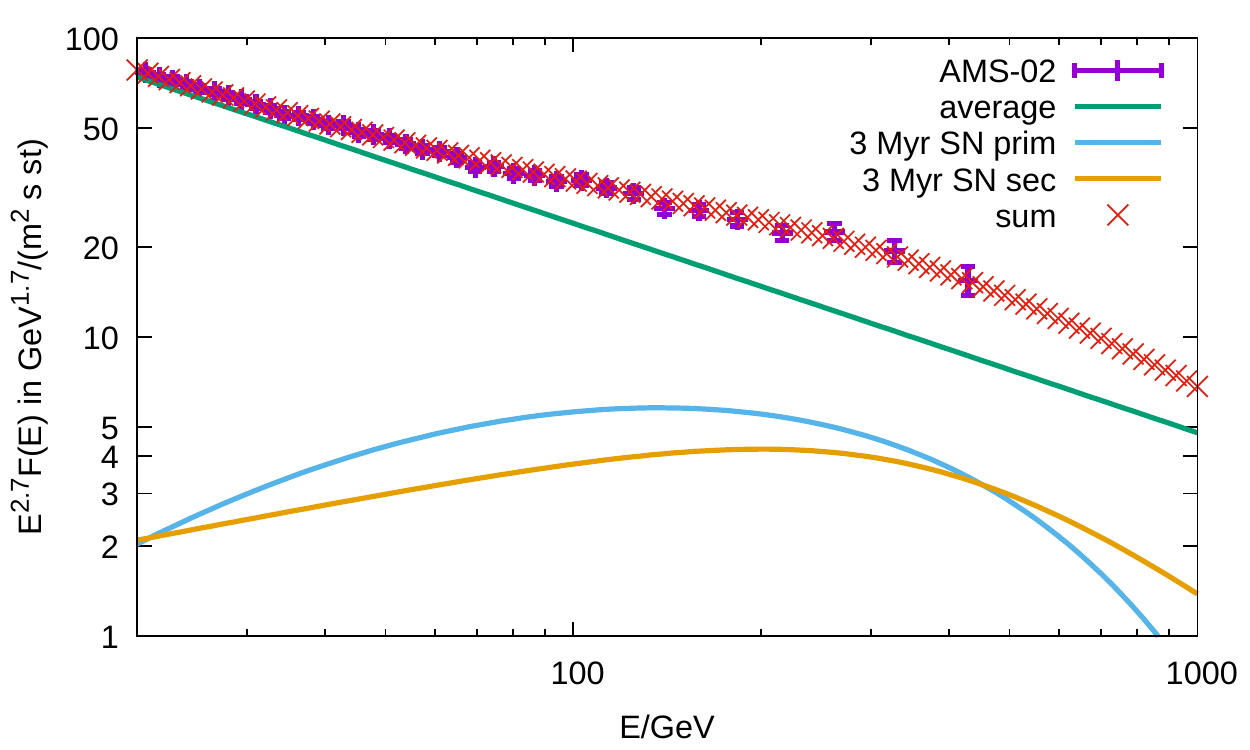}
\caption{Electron fluxes measured by AMS-02 compared to the 
contributions for the local source (red/rose solid curves) and average (dashed black/grey) components. 
\label{fig:ele}}
\end{figure}

We discuss now the electron fluxes expected in our model. Similar to CR
nuclei, electrons are injected as primaries in the acceleration process.
Neglecting energy losses, the contribution for the electron spectrum from
the local source has therefore the same functional form as the proton
spectrum and can be expressed as $C^{(2)}_{e^-}F^{(2)}(\R)$. The normalisation
constant $C^{(2)}_{e^-}$ is related to the only poorly restricted
electron/proton ratio $K_{\rm ep}$ at injection which we choose as
$K_{\rm ep}=4\times 10^{-3}$.  The energy losses via synchrotron and inverse
Compton emission lead to an exponential cutoff for primary electrons,
since they are all injected instantaneously 2.7\,Myr ago. From the data in
the positron spectrum, we choose the  break at $E_{\rm br}=300$\,GeV. The
spectrum of primary electrons from the local source is therefore modified as
$$
C^{(2)}_{e^-}F^{(2)}(\R) \to  C^{(2)}_{e^-}F^{(2)}(\R) \exp(-E/E_{\rm br}) \,.
$$
The low-energy component $F^{(1)}(\R)$ representing
the average CR flux from sources in the disk is modified by energy losses as
$F^{(1)}(\R) \to  F^{(1)}(\R)(E/E_0)^{0.5}$ \cite{panov}.
In Fig.~\ref{fig:ele}, we compare the resulting electron intensity compared
to the data from AMS-02. In addition, we show also the flux of secondary
electrons which are continuously produced by proton-gas interactions. This
flux coincides with the the secondary flux of positrons from the local source
discussed earlier, and becomes dominant above 500\,GeV.

\section{Cosmic ray secondary nuclei and the B/C ratio}

Apart from positrons and antiprotons, the propagation of CRs injected by 
the local source results in the production of secondary cosmic ray nuclei. 
In particular, the spallation of primary carbon and oxygen
generates a secondary boron flux. Similarly to the flux of electrons and 
positrons, the flux of boron nuclei is expected to have two contributions: 
one generated by the average cosmic ray flux and the second one produced 
by the carbon injected by the local recent supernova.  

Contrary to the positron and antiproton fluxes, the local source contribution 
to the boron flux is expected to appear around TV, rather than 100\,GV.
This is because the typical rigidity of positrons and antiprotons produced in 
CR interactions is below $10\%$ of the rigidity of the primary cosmic ray, 
while the rigidity of boron nuclei produced in spallation of the carbon nuclei 
is identical or very similar to that of the primary carbon. 

We adopt a simple leaky-box model approach, 
\be
\frac{F_B}{F_C}\simeq (X/20\mbox{ g/cm}^2)/\left(1+X/(13\mbox{ g/cm}^2)\right)
\ee
using as parameters those determined in Ref.~\cite{Webber03,Blum13}:
The characteristic column densities 20\,g/cm$^2$ and 13\,g/cm$^2$ correspond 
to the grammage on which a significant part of the carbon is transformed into 
boron and significant part of boron is destroyed by spallation, respectively. 
Within the considered model the total carbon flux is $F_C=F^{(1)}_C+F^{(2)}_C$. 

In both isotropic and anisotropic diffusion models the residence time of CRs 
in the interstellar medium scales as a power law of particle rigidity 
$T\propto \R^{-\beta}$ in the energy range of interest. The column density 
traversed by the CRs belonging to the average flux component scales as 
$X^{(1)}=n_{\rm ISM}T\propto \R^{-\beta}$
so that the expected rigidity dependence of the B/C ratio is a power-law, 
$F_B/F_C\propto \R^{-\beta}$ for $F_B/F_C\ll 0.7$. At the same time, carbon 
injected by the local source has traversed the fixed column density 
$\simeq (1.0-1.4)$\,g/cm$^2$ since the moment of injection, which is
determined by our fit to the positron spectrum in the previous section.
The overall amount of boron contained in the CR flux is, therefore
\begin{eqnarray}
\frac{F_B}{F_C}&=&\frac{X^{(1)}/(20\mbox{ g/cm}^2)}{\left(1+X^{(1)}/(13\mbox{ g/cm}^2)\right)}\frac{F^{(1)}_C}{F_C}
\nonumber
\\ && +\frac{X^{(2)}/(20\mbox{ g/cm}^2)}{\left(1+X^{(2)}/(13\mbox{ g/cm}^2)\right)}\frac{F^{(2)}_C}{F_C} .
\end{eqnarray}
This flux ratio is shown in Fig.~\ref{fig:bc} together with the measurements 
from AMS-02~\cite{AMS02_BC}.  The thin dashed black and grey lines 
show the rigidity scaling of the B/C ratio which would be found in the 
absence of the local source component. We have assumed the power-law slope 
$\beta=1/3$ for the rigidity dependence of the grammage, appropriate for 
Kolmogorov 
turbulence.  Note that the slope of the B/C ratio  disagrees at low rigidities 
with the measurements. The red and rose  horizontal lines correspond to 
the B/C ratio which would be found if the observed flux would be fully provided
by the local source for the grammage $1.2\pm0.6$\,g/cm$^2$ (the 50\% uncertainty range takes into account the uncertainty of the estimate based on positron flux from the local source). The thick dotted black and grey lines correspond to the case of 
the two-component carbon flux shown in Fig. \ref{fig:nuclei}. Above few 
$\times 100$\,GV, the B/C 
ratio flattens  and reaches finally a plateau corresponding 
to the B/C ratio produced by the local source.

The plateau in the B/C ratio and an approximately constant antiproton-proton
ratio predicted in the local source model distinguish this model
from the ``re-acceleration model'' which, according to
Refs.~\cite{reac,reac2}, predicts rising  antiproton-proton and B/C ratios. 

In our analysis we have considered a model with a single local source 
superimposed onto a smooth "sea" of CRs. In reality, the transition between 
the rigidity range where the assumption of continuous CR is justified
and discrete sources is gradual. This should be reflected in the rigidity 
dependence of the B/C ratio: several plateau-like features might be present, 
so that the overall plot should have a set of "steps" corresponding to 
plateaus connected to several recent injection episodes.

\section{Detectability of the local source signature in the B/C ratio}

Step-like features in the energy dependence of the B/C ratio and in particular
a plateau ${\rm B/C}\sim {\rm few} \times 0.01$ above TV are specific 
predictions of 
the local source model. The detection of such a plateau or other step-like
features would provide a powerful test of the model. This test would be 
complementary to the test of another prediction of the model: a simple 
power-law extension of  the antiproton flux into the TeV band from lower
energies, i.e.\ no softening of the spectrum, cf.\ with 
Fig.~\ref{fig:pos_aprot}. The theoretical predictions for the antiproton flux 
suffer from  uncertainties in the production cross-section, 
of the average primary CR spectrum and of the propagation of CRs through the 
interstellar medium and magnetic fields. In contrast, the theoretical 
predictions for the B/C ratio expected from the average CR flux are not 
sensitive to the details of the CR spectra in the interstellar medium. They are largely determined by the energy dependent grammage traversed by the carbon and oxygen nuclei during their residence time in the interstellar medium. 

\begin{figure}
\includegraphics[width=\columnwidth]{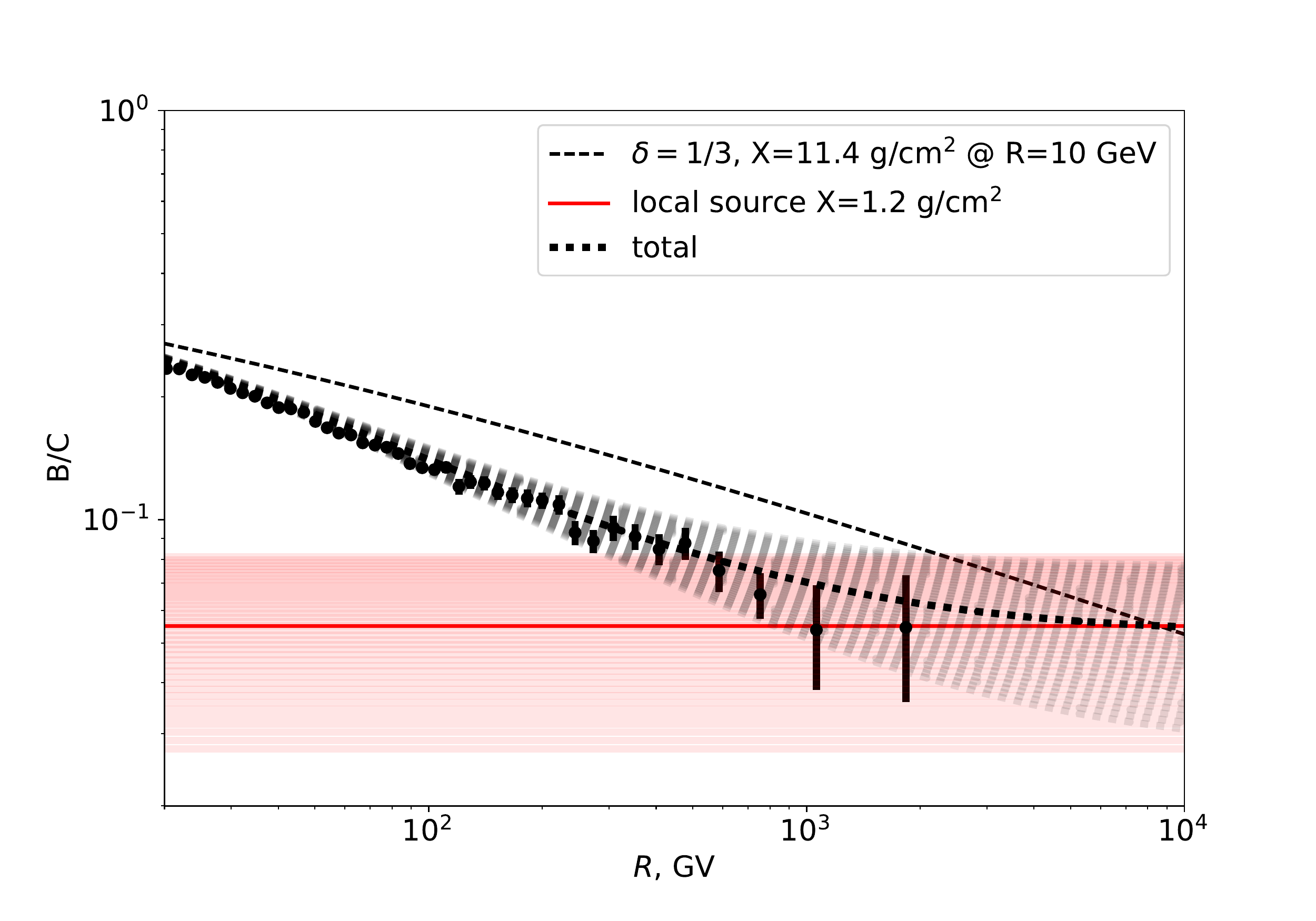}
\caption{Rigidity dependence of B/C ratio of the average cosmic ray flux  in a model with diffusion coefficient determined by the Kolmogorov turbulence spectrum of magnetic field (thin dashed line), by a fixed grammage traversed by the carbon nuclei originating from a several Myr old local supernova (red solid line and and shaded uncertainty band) and in a model in which the observed cosmic ray flux is a sum of average and local source contributions in proportions deduced from the template fit to the carbon spectrum (black dotted line and grey uncertainty band).   }
\label{fig:bc}
\end{figure}

The energy range of the measurement of the B/C ratio by AMS-02 (shown in 
Fig.~\ref{fig:bc}) only touches the predicted "plateau" energy range. 
AMS-02 measurements are intrinsically restricted to the rigidity range below 
10\,TV because of the limited charge resolution at higher rigidities. 
Dedicated detectors optimized for the measurement of elemental composition 
of the CR flux in the multi-TV range are needed for the detection of 
the local source related plateau of the B/C ratio. 

Measurements by the CREAM balloon born detector also do not extend to the 
rigidity range of interest. The carbon nuclei flux is at the level of 
$F_C=1.5_{-0.6}^{+1.0}\times 10^{-9}$ 1/(m$^2$ sr s GeV/n) at the energy per 
nucleon $E/A=7.4$~TeV (corresponding to the rigidity $\R=cp/Z\simeq 15$~TV).
A CREAM-type detector accumulates the 
carbon signal at the rate of 0.2 events per day at this energy/rigidity. 
Only five carbon nuclei were detected in the first CREAM flight in this energy 
range. No boron nuclei are detectable in a flight campaign if the B/C ratio 
is at the expected plateau level, $F_B\sim 0.06 F_C$.

Larger event statistics could potentially be accumulated in the case of a 
multi-year exposure with a space-based detector, like ISS-CREAM or NUCLEON.  
A nominal five-year CREAM-like space detector could yield the event statistics
  of about 215 events in the energy range of interest.  The expected 
statistics of the boron cosmic rays is about 12~events accumulated over the 
mission time span. This shows that measurement of the flattening of the 
energy dependence of the B/C ratio would be challenging with  planned or 
operating space-based CR detectors. Larger detectors and/or longer 
exposure times are desirable for the detection of the predicted plateau. 

Another prediction of
our model are step-like features in the  B/C ratio: the
``average'' flux is the sum of $N\simeq 10$--20 local SNe, each of it 
producing its own plateau. Summing up these contribution will produce
step-like feature in the B/C ratio, with the plotted 1/3 power-law
emerging only in the $N\to\infty$ limit. These features might be present 
at lower energies and would be more readily detectable increasing 
the statistics of the B/C signal.

\section{Discussion and conclusions}

We have shown that the whole set of known peculiar features ($i$)-($vi$) 
of the locally observed CR spectrum can be explained within a single 
self-consistent model. This model takes into account the contribution to the 
CR flux of a supernova which has injected CRs within a distance of about 
100\,pc from the Sun some 2--3 Myr ago. It is likely that this supernova is
connected to the deposition of $^{60}$Fe isotopes in the deep ocean crust 
of the Earth.

In Ref.~\cite{savchenko}, we showed that a CR source with age $T=2$\,Myr
and distance $d=200$\,pc leads to a dipole anisotropy 
$\delta \simeq 5\times 10^{-4}$, assuming quasi Gaussian propagation of CRs.
This value corresponds to the central value of the observed
range for $\delta$. For the source distance and age used in the present work, 
the  dipole anistropy  $\delta \simeq 2\times 10^{-4}$ follows,
which corresponds to the values at the lower edge of the error band shown in 
Fig.~2 of  Ref.~\cite{savchenko}. As a possible reason for this small 
deviation we note that the local source may reside inside  the complex of
the local and Loop~1 superbubble. As a result, the bubble wall between Loop~1 
and the local superbubble might act as ``effective'' CR source, leading to 
deviations from the naive dipole formula.

Figure~\ref{fig:nuclei} shows the explanation for the features ($i$) and 
($ii$). The difference in the spectral slopes of different nuclear species 
in the CR flux is explained by the slightly different normalisation of the 
local source contribution to the flux of different nuclei (related to the
particular chemical composition of the SN environment) and due to the 
rigidity dependent shifts of the local source component along the energy axis. 
The breaks in the spectra of individual nuclei at several hundred GeV are 
due to the fact that the local source component gradually starts to dominate 
over the average cosmic ray flux component at higher energies. 

The measurement of the cooling break in the spectrum of CR positrons at 
300\,GeV (cf.\ with Fig.~\ref{fig:pos_aprot}) determines the 
time elapsed since the episode of CR injection by the supernova. This time 
is consistent with the age of the supernova deduced from the measurements of 
the $^{60}$Fe deposition in the deep ocean crust. The presence of the 
positrons and antiprotons produced in interactions of the CRs injected by 
the supernova explains the features ($iii$)--($iv$). 

Low energy cosmic rays from the supernova have not yet reached the ordered 
magnetic field line passing through the Solar system. The low-energy 
suppression of the local source component explains the energy dependence 
of the dipole anisotropy ($vi$). This suppression is also evident from the 
phenomenological fit of the shape of the local source contribution as an 
excess flux over the average flux of CRs in the local Galaxy, as deduced 
from the \gr\ observations of nearby molecular clouds. 

The model discussed above is over-constrained which allows one to falsify it. 
Four testable predictions of the model are the persistence of a 
nearly constant slope of the antiproton spectrum  in the energy range beyond 
300\,GeV, the positron-electron ratio $R$, step-like features and the 
flattening of the boron-to-carbon 
ratio to a "plateau" level of $F_B/F_C\sim 0.06$ in the 1--10\,TV range. 
These tests are complementary. The test based on the antiproton-proton 
and the positron-antiproton ratio suffers from uncertainties in theoretical  
calculation of propagation of secondary particles through Galactic magnetic field, as illustrated in section III. The tests based on the 
measurement of B/C ratio are challenging from the experimental point of 
view. The detection of the plateau requires a decade-scale exposure with 
space-based cosmic ray detectors like ISS-CREAM and NUCLEON. 
While the exact numerical values of these predictions is influenced
in particular by the modelling of the local magnetic structure and
the position of the local source, the presence of these features depends 
only on the dominance of a local source in the CR flux in the 
$\sim 1-30$\,TeV range.

We also note that a supernova as close as 100\,pc may also impact
the Earth's atmosphere and biota, and the absence of a major extinction
2--3\,Myr ago is another evidence for the low-energy suppression
of the CR flux from the local source~\cite{bio}.

Finally, we want to stress the differences between our scenario and related
models. In Ref.~\cite{Fujita:2009wk}, the effect of one or several nearby SN
explosions on the secondary ratios was considered. Since these authors
assumed standard isotropic diffusion, both the primary and the secondary
fluxes of these sources in general do not dominate the measured fluxes today.
In particular, the scenario of  Ref.~\cite{Fujita:2009wk} can therefore not
explain the plateau in the dipole anisotropy or the breaks in the primary
spectra. In order to enhance the secondary fluxes, the SN explosions had to
occur in a dense environment, as e.g.\ a molecular cloud. 
The authors of Ref.~\cite{reac2} also assumed standard isotropic
diffusion and split the contribution of all SNRs into a young and an old
component. Their main prediction is a featureless B/C ratio, in contrast
to the step-like features typical for our model.

\acknowledgments
This research was supported in part with computational resources at NTNU 
provided by NOTUR, \url{http://www.sigma2.no}. We would like to thank the
anonymous referee for valulable comments which helped to improve our 
manuscript.

\appendix

\section{Calculation of trajectories}

In the standard approach to cosmic ray (CR) propagation, their time evolution
phase space is approximated as adiffusion process. Inncluding CR interactions 
in the resulting transport equation allows one to calculate
self-consistently fluxes of primary and secondary CRs. However, this approach
has several disadvantages: Most importantly, the diffusion approximation
breaks down in several regimes~\cite{GKS14,1,2}. Moreover, the diffusion tensor
is an external input which can be only loosely connected to fundamental 
properties of the Galactic magnetic field (GMF). Therefore we  calculate 
the path of individual CRs solving the equations of motion of particles 
propagating in the GMF. In particular, we employ the code described and 
tested in Ref.~\cite{0}  which uses nested grids. Our procedure to use 
such grids is the following:  For the largest scale we choose $L_{\max}=25$\,pc 
according to the recent LOFAR measurement of the maximal scale of magnetic 
field fluctuations in the disk~\cite{LOFAR}. Then we use up to four nested 
grids which cover five decades in scale such that the  smallest resolved
length scale is a factor 10 below the resonannce scale for a CR with $10$\,TV
rigidity. This corresponds to the lowest rigidity for which  we made direct 
calculations in Ref.~\cite{PRL}.

For the calculation of the local CR flux, we have divided the nearby 
Galaxy in cells and saved the length of CR trajectories per cell.
Compared to Ref.~\cite{PRL}, we increased the number of cells and observers:
We  divided the Galactic plane into non-uniform grid with cell size 
$20\,{\rm pc}\times 20$\,pc at the position of the SN, which gradually 
increase to $100\,{\rm pc}\times 100$\,pc at distances  more than 500\,pc.
The vertical height of these cells was chosen as 20\,pc. 
This allowed us to calculate simultaneously the flux at various places
with differing perpendicular distance to the regular magnetic field line 
which goes through the SN (see Fig. 3).

Several detailed models for the regular component of the GMF exist
and two of the most recent and detailed ones are the models of
Jansson-Farrar~\cite{JF} and of Pshirkov et al.~\cite{Ps}.
Reference~\cite{0} compared both models and found that they lead
qualitatively to the same predictions for CR propagation. 
The turbulent part of the magnetic field can be characterized by the 
power-spectrum $\mathcal{P}(\vec k)$ and the correlation $L_{\rm c}$ of 
its fluctuations. Assuming a power-law $\mathcal{P}(k)\propto k^{-\alpha}$
for the spectrum,  the maximal length  $L_{\max}$ of the 
fluctuations and the correlation length $L_{\rm c}$ are connected by 
$l_{\rm c} = (\alpha-1) L_{\max}/ (2\alpha)$.
An important constraint on CR propagation models comes from ratios of 
stable primaries and secondaries produced by CR interactions on gas 
in the Galactic disk.  Employing the 
Jansson-Farrar  model, Ref.~\cite{0} found that the  B/C ratio 
can be reproduced choosing as the maximal length of the fluctuations 
$L_{\max} = 25$\,pc and $\alpha=5/3$, if the turbulent field is rescaled 
to one tenth of the value proposed by Jansson-Farrar~\cite{JF}.
For this choice of parameters, CR propagation in the Jansson-Farrar 
model reproduces a large set of local CR measurements and we used these
parameters in this work. In particular, the B/C ratio measured by
AMS-02~\cite{AMS02_BC} can be nicely reproduced using a Kolmogoroff power
spectrum, $\alpha=5/3$, as we have shown here.

Cosmic rays with energies  $E\lsim 100$\,TeV propagate in the diffusive 
regime, using our standard magnetic field model. In the energy range 
10--100\,TeV, we showed that the diffusion is self-similar, i.e.\ the 
propagation at $E<E_0$ follows the relation $t' = t (E_0/E)^{1/3}$, see 
Fig.~4 in Ref.~\cite{PRL}. Here we applied this self-similar solution to 
all energies lower than $E_0=10^{14}$\,eV.


\end{document}